# Optimization of a Commercial Injection-Moulded component by Using DOE and Simulation

Mandana Kariminejad, Centre for Precision Engineering, Material and Manufacturing, Institute of Technology Sligo
David Tormey, Centre for Precision Engineering, Material and Manufacturing, Institute of Technology Sligo
Saif Huq, School of Computing and Digital Media, London Metropolitan University
Jim Morrison, Department of Electronics and Mechanical Engineering, Letterkenny Institute of Technology
Jeff Redmond, Combination Products, Science and Technology, AbbVie Inc.
Carlos Souto, Engineering Moulding, AbbVie Ballytivnan
Marion McAfee, Centre for Precision Engineering, Material and Manufacturing, Institute of Technology Sligo

*Abstract*

Injection moulding is an important industry, providing a significant percentage of the demand for plastic products throughout the world. The process consists of many variables which directly or indirectly influence the part quality and cycle time. The first step in optimizing the process parameters is identifying the most significant variables affecting the desired output. For this purpose, various Design of Experiments methods (DOE) have been developed to investigate the effect of the experimental variables on the output and predict the required settings to achieve the optimal value of the output. In this study we investigate the application of DOE for a commercial injection moulded component which suffers from a long cycle time and high shrinkage. The Taguchi method has been used to analyze the effect of four input variables on the two output variables: cycle time and shrinkage. The component has been simulated in the Moldflow software to validate the predicted output and optimized settings of the variables from the DOE. Comparison of the simulation result and the predicted value from the DOE illustrated good accordance. The calculated optimal setting with the Taguchi method reduced the cycle time from the 40s to about 23s and met the shrinkage criteria for this commercial part.

## 1. INTRODUCTION

One of the most developed processes for the production of plastic components is injection moulding. In general, this process contains three main steps: the filling stage in which melted polymer pellets are injected into the cavity, the packing stage which prevents excessive shrinkage by injection of extra polymer, and the cooling stage where the polymer solidifies and gets ready for ejection (Kazmer, 2007). During these stages, many process parameters such as mould temperature, melt temperature, and injection pressure should be controlled and adjusted, directly affecting the part quality and efficiency of the process. Non-optimal process settings not only lead to defects in injection moulded parts such as warpage, shrinkage and residual stresses, but also cause long cycle time and low process efficiency (Kim et al., 2009; Xu et al., 2015; Zhang & Jiang, 2007).

The first step for improving quality and enhancing efficiency is to identify the most significant process parameters influencing the quality factors. For this purpose, various Design of Experiment (DOE) methods have been developed. One of the developed DOE methods for prediction, optimization, and selection of the key variables is the Taguchi method. The main advantage of this method is designing the experiments based on an orthogonal array with a minimum number of experiments which saves time and cost (Van Nostrand, 2002). This method has been used in injection moulding for optimization of the process in various studies. Ozcelic and Erzurumlu (Ozcelik & Erzurumlu, 2006) investigated the effect of seven factors on the warpage of thin shell plastic components using the Taguchi method and specified the key parameters influencing the warpage. Zhang et al. (Zhang & Jiang, 2007) first used a fractional factorial design to identify the main factors on the part quality and then used Taguchi method to optimize these process factors. Altan (Altan, 2010) investigated the impact of different process parameters on the shrinkage of polypropylene (PP) and polystyrene (PS) injection moulded parts using Taguchi method and ANOVA. They concluded that the most significant factor in the shrinkage is packing pressure for PP and melt temperature for the PS. Then a neural network based method was applied to predict shrinkage for these two parts based on the optimal process levels from the Taguchi result. Jan et al. (Jan et al., 2016) applied Taguchi method and response surface method to predict sink marks in the injection moulding process. Moayyedian et al. (Moayyedian et al., 2018) used a combination of Taguchi method and fuzzy logic to

Email: Mandana Kariminejad
  Mandana.kariminejad@mail.itsligo.ie

optimize three key defects: shrinkage, warpage and short shot, in injection moulding. Hentati et al. (Hentati et al., 2019) studied the effect of four process parameters on the shear stress in PC/ABS blended part and validated the result by simulation in SOLIDWORKS software.

The optimization of the cycle time and shrinkage of a commercially moulded component from an industrial partner, AbbVie, is studied and presented in this paper. For this purpose, the effect of four input factors, melt temperature, mould temperature, injection pressure, and holding time, has been studied with respect to two critical outputs: cycle time and shrinkage for this product.

## 2. METHODOLOGY

### 2.1 Part description

In this study, we investigate a component which we refer to as a 'clip'. The isometric view of the clip is illustrated in Figure 1. The initial process setting for optimization has been provided by AbbVie Ballytivnan, Sligo. The material of the Clip component is Delrin 500P NC010 and the dimension is 32.36×26.33×11.9 mm.

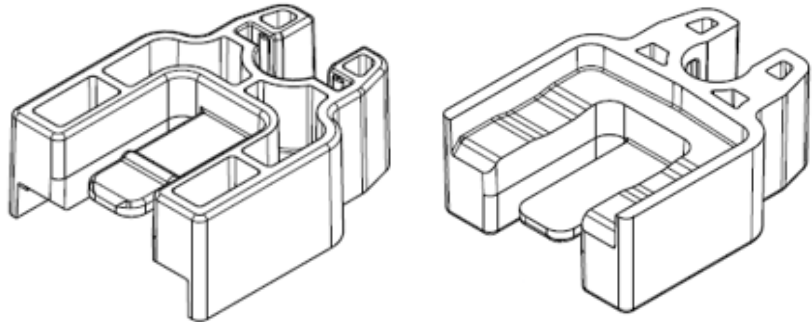

*Figure 1. Isometric view of the Clip injection moulded component*

### 2.2. Simulation

Autodesk Moldflow Insight 2019 software has been used to simulate the injection moulding process and validate the data from DOE for the Clip. The simulated part with the designed cooling channels and two cavities and two injection locations has been shown in Figure 2. (a). The conventional cooling channels (blue channels) with two baffles (yellow channels) at the middle of cooling circuits have been indicated in Figure 2. (a). The baffle is a type of cooling channel with a blade at the centre, placed at the hot spots, which causes an increase in the turbulency and heat transfer, thus a reduction in the cooling time. Figure 2. (b) shows the simulated component with immobile and mobile moulds and ejector pin spots. For the finite element analysis, the Dual-domain mesh (fusion) has been selected because of the part geometry and the mesh tool has been applied to eliminate the mesh defects.

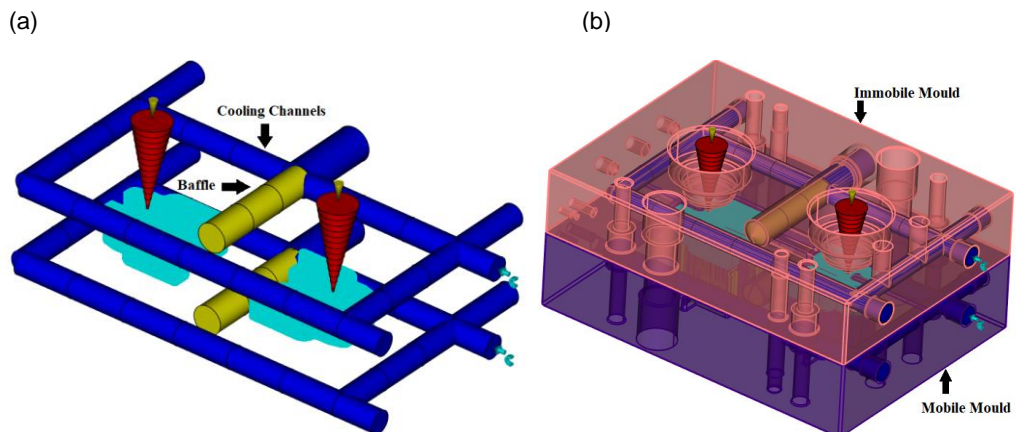

*Figure 2. **(a)** Simulated Clip part with the designed cooling channels. **(b)** The Clip with mould and cavity.*

In this study, for the initial optimization of the process and saving cost and time, instead of running the designed experiments from Taguchi in the real process, each experiment has been run in the simulation. For examining the trustworthiness of the simulation, the result of a specific injection moulding process setting has been compared to the simulation in Moldflow. The result of this comparison has been summarized in Table 1. Figure 3 indicates the result from the simulation for the filling time. The relatively small error percentage between the simulation and the actual process demonstrates that the simulation can be used for initial optimization instead of the real experiment.

Table 1. Comparison of the real process and simulation

| Parameters | Real Process | Moldflow Simulation | Error % |
|---|---|---|---|
| Cycle time (s) | 40 - 46 | 43.02 | 6.9 |
| Filling time (s) | 0.355 | 0.37 | 4.05 |
| Cooling time (s) | 30 | 28.03 | 7.02 |

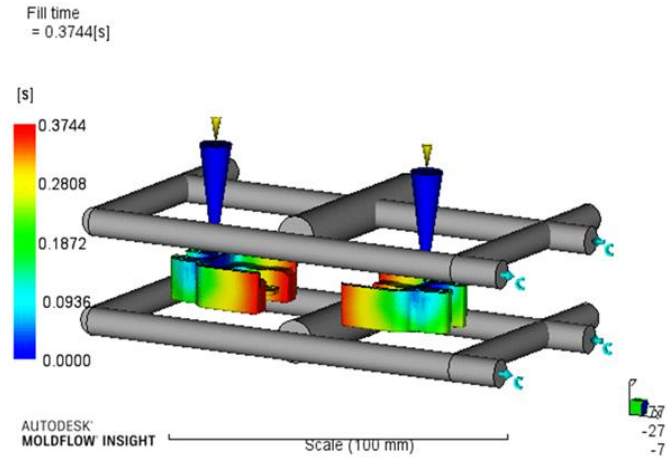

Figure 3. The result of filling time from Moldflow simulation

## 2.3. Taguchi method

Taguchi method is a type of Design of Experiments method that can be used not only for the screening of variables, but also for optimization. This method is a combination of fractional factorial design and orthogonal array. The orthogonal experimental setting in this method refers to an equivalent number of all levels for each variable in the designed experiments, ensuring the balance of the array (Butler, 1992; Kr Dwiwedi et al., 2015; Van Nostrand, 2002).

This method has been used in this study to investigate the effect of injection moulding process parameters on the part shrinkage and cycle time. Each of the input factors has three levels based on the primary process setting from the industrial partner. Minitab 19 software has been used to find the optimal process parameters via the Taguchi method. The detailed description of the input parameters has been summarized in Table 2.

Table 2. Input process Parameters details

| Input Parameters | Level 1 | Level 2 | Level 3 |
|---|---|---|---|
| Mould temperature (°C) | 75 | 80 | 85 |
| Melt temperature (°C) | 215 | 220 | 230 |
| Injection pressure (bar) | 470 | 530 | 580 |
| Holding time (s) | 3.5 | 4.5 | 5.5 |

The L9 orthogonal array has been used based on the Taguchi method shown in Table 3. The optimal output ($R_{opt}$) can be calculated from equation (1) for four input variables ($A$, $B$, $C$, and $D$). $\bar{R}$ is the average of all outputs from nine experiments and $\bar{A}_x, \bar{B}_x, \bar{C}_x \text{ and } \bar{D}_x$ are the average of the desired output at the optimum level of $x$. As it is clear from Table 3, the number of experiments for four input variables and three-levels is just nine with the Taguchi method, while for the full factorial design, this number would increase to $3^4 = 81$.

$$R_{opt} = \bar{R} + (\bar{A}_x - \bar{R}) + (\bar{B}_x - \bar{R}) + (\bar{C}_x - \bar{R}) + (\bar{D}_x - \bar{R}) \tag{1-a}$$

$$\bar{R} = \frac{R_1 + R_2 + R_3 + \cdots + R_9}{9} \tag{1-b}$$

Table 3. L$_9$ orthogonal array Taguchi method

| No. | Mould Temperature(°C) | Melt Temperature (°C) | Injection Pressure (bar) | Holding time (s) |
|---|---|---|---|---|
| 1 | 75 | 215 | 470 | 3.5 |
| 2 | 75 | 220 | 530 | 4.5 |
| 3 | 75 | 230 | 580 | 5.5 |
| 4 | 80 | 215 | 530 | 5.5 |
| 5 | 80 | 220 | 580 | 3.5 |
| 6 | 80 | 230 | 470 | 4.5 |
| 7 | 85 | 215 | 580 | 4.5 |
| 8 | 85 | 220 | 470 | 5.5 |
| 9 | 85 | 230 | 530 | 3.5 |

The signal-to-noise ratio is a quality indicator to evaluate the variation of a specific variable on the final output (Ross PJ., 1996). In the injection moulding process, the aim is to minimize the cycle time and shrinkage as much as possible. Hence, in this study, the Taguchi signal-to-noise ratio $S/N$ should be defined as *''the-smaller- the-better''* described in Equation 2. '*n*' is the number of experiments (here 9), and '$y_i$' is the response value for the *i*th experiment.

$$S/N = -10 \log_{10} \left( \frac{\sum_{i=1}^{n} y_i^2}{n} \right) \tag{2}$$

## 3. RESULTS AND DISSCUSSION

The designed experiments based on Table 3 have been simulated in the Moldflow software and the result for cycle time and shrinkage and the related signal-to-noise ratio have been summarized in Table 4.

The cycle time in this simulation is made up of the filling time, packing time, cooling time, and mould open time. For the shrinkage simulation, first, the critical dimensions and the related tolerances provided by AbbVie are defined. The shrinkage has been examined based on the average linear shrinkage, that is, the equally-weighted mean of parallel and perpendicular shrinkage. The nominal parallel and perpendicular shrinkage is 1.934% and 2.082% for Delrin 500P NC010, respectively. The shrinkage result should be below these nominal values to prevent excessive shrinkage in part.

Table 4. Simulation result for L9 orthogonal array

| No. | Mould Temperature(°C) | Melt Temperature (°C) | Injection Pressure (MPa) | Holding time (s) | Cycle time (s) | Shrinkage (%) | S/N Cycle time | S/N shrinkage |
|---|---|---|---|---|---|---|---|---|
| 1 | 75 | 215 | 47 | 3.5 | 49.4161 | 2.2 | -33.87 | -6.84 |
| 2 | 75 | 220 | 53 | 4.5 | 51.0519 | 2.183 | -34.16 | -6.78 |
| 3 | 75 | 230 | 58 | 5.5 | 54.4495 | 2.571 | -34.7 | -8.2 |
| 4 | 80 | 215 | 53 | 5.5 | 29.3798 | 1.992 | -29.36 | -5.98 |
| 5 | 80 | 220 | 58 | 3.5 | 30.4038 | 2.093 | -29.65 | -6.41 |
| 6 | 80 | 230 | 47 | 4.5 | 32.3585 | 2.062 | -30.1 | -6.28 |
| 7 | 85 | 215 | 58 | 4.5 | 22.925 | 1.972 | -27.2 | -5.89 |
| 8 | 85 | 220 | 47 | 5.5 | 23.4541 | 1.961 | -27.4 | -5.84 |
| 9 | 85 | 230 | 53 | 3.5 | 24.4298 | 2.144 | -27.75 | -6.62 |

### 3.1 Screening of input parameters

The Taguchi method is able to assess the most effective level and the importance rate of each input variable on the desired output. The result of average values for cycle time and shrinkage has been summarized in Figure 4.

Regarding Figure 4. (a), the most significant factor on cycle time is mould temperature (Tmold). The minimum value of cycle time will be obtained if the mould temperature is set to the highest level (85°C). Melt temperature (Tmelt), holding time (tholding) and injection pressure (Pinj) also affect cycle time in that order of importance; however, their influence is not considerable.

Figure 4. (b) indicates mould temperature is also the leading variable affecting shrinkage, and to minimize the shrinkage, the mould temperature should be set at the highest level of 85°C. The influence of melt temperature is almost major and for the optimization of shrinkage, the minimum level of 215 °C should be adjusted. The holding time and injection pressure have similar effects on the linear shrinkage. Injection pressure should be fixed at the minimum level (47 MPa) and holding time should be set at the medium level, which is 4.5 s. The importance of each input variable on the outputs has been presented in Table 5, where the input with the highest and lowest impact has been defined by Rank '1' and Rank '4', respectively.

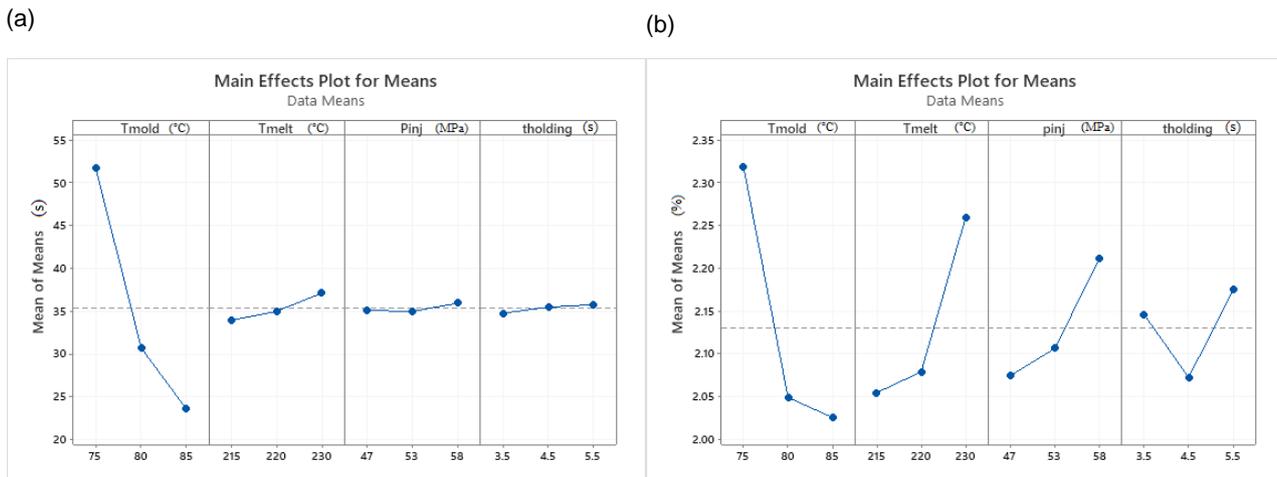

*Figure 4. Average values plot for **(a)** cycle time, **(b)** Shrinkage at three levels*

*Table 5. The effect of each input variables on the desired outputs*

| Desired Outputs | Mould Temperature(°C) | Melt Temperature(°C) | Injection Pressure (MPa) | Holding Time(s) |
|---|---|---|---|---|
| **Cycle Time(s)** | 1 | 2 | 4 | 3 |
| **Shrinkage%** | 1 | 2 | 3 | 4 |

*3.2 Optimization of outputs with Taguchi method and simulation*

The Taguchi method estimates the optimum output based on the optimal setting from screening in section 3.1 by Equation 1. For validation of the predicted values from the Taguchi method, the predicted optimal settings were simulated in Moldflow. As shown in Table 6, the difference between the prediction from the Taguchi method and the Moldflow simulation is below 10% which validates that the Taguchi method can successfully predict optimal settings. The shrinkage percentage is below the nominal value of the Delrin 500P NC010, which verifies that under this process setting, excessive shrinkage will not occur in the part. Obviously the simulation should be followed by optimisation of the settings in the actual process, however based on the Taguchi method (Table 6) applied to the simulation environment, the initial mould temperature should be fixed at the highest level and the initial melt temperature at the lowest level. Besides that with this optimal setting, the cyle time declined from almost 40 s to 23s, improving the process efficiency

*Table 6. Comparison of the outputs from Taguchi method and Moldflow simulation*

| Output Parameters | Mould Temperature (°C) | Melt Temperature (°C) | Injection Pressure (MPa) | Holding Time (s) | Taguchi Predicted Value | Moldflow Simulation Value | Error % |
|---|---|---|---|---|---|---|---|
| Cycle Time(s) | 85 | 215 | 53 | 3.5 | 21.2575 | 22.92 | 7.27 |
| Shrinkage% | 85 | 215 | 47 | 4.5 | 1.83 | 1.98 | 7.57 |

## 4. CONCLUSION

In this paper, Taguchi method and simulation are applied together to study the effect of melt temperature, mould temperature, packing temperature and holding time on the shrinkage and cycle time of the commercial injection moulded part. The experiments were initially simulated in the Moldflow software instead of the actual process to save time and cost.

The most significant factor on both shrinkage and cycle time is mould temperature. The result indicated that 85°C of mould temperature, 215°C of melt temperature, 53 Mpa of injection pressure, and 3.5 s of holding time minimize the cycle time to almost 23 s, much less than the current cycle time of the part in the process which is about 40 s. The simulation obtained a minimum shrinkage of 1.98% with a mould temperature of 85°C, melt temperature of 215°C, injection pressure of 47 Mpa, and 4.5 s of holding time (See Table 6). This value is lower than the nominal shrinkage of the material (nominal parallel and perpendicular shrinkage are 1.934% and 2.082%). Based on this study, the mould temperature should be set at the highest level and melt temperature at the lowest level to optimize shrinkage and cycle time. Changing the injection pressure and holding time is not significant on the cycle time, so they should be fixed at the minimum and middle levels for minimum shrinkage, respectively.

Further research to improve the optimization results includes validation of the simulation data by running the L9 in the real injection moulding process, increasing the number of experiments from L9 to L27 to investigate the interactions between the factors and study other input variables such as ejection temperature, flow rate, coolant temperature, gate type and cooling channels on the shrinkage and cycle time.